\documentstyle[twoside]{article}

\def\npb{\nopagebreak}
\def\nn{\nonumber\\}
\def\nq{\hspace{-1em}}
\def\nqq{\hspace{-2em}}
\def\nhq{\hspace{-0.5em}}

\def\cm{\hspace{1cm}}

\def\nx{\vspace{-1ex}}
\def\nxx{\vspace{-2ex}}
\def\nhx{\vspace{-0.5ex}}
\def\al{&\nhq}
\def\noi{\noindent}

\def\beq{\begin{equation}}
\def\eeq{\end{equation}}
\def\bear{\begin{eqnarray}}
\def\ear{\end{eqnarray}}

\newcommand{\Acknow}[1]{\bigskip{\bf Acknowledgement}\\[1ex] \noindent #1}

\newcommand{\bls}[1]{\renewcommand{\baselinestretch}{#1}}

\def\sign{{\rm sign}\,}
\def\const{{\rm const}}

\def\e{{\rm e}}
\def\to{\rightarrow}
\def\chg{\leftrightarrow}
\def\gg{\overline{g}}
\def\pp{\overline{p}}
\def\qq{\overline{q}}
\def\g{\hat g}
\def\ds{d{\hat s}^2_D}
\def\half{{\textstyle\frac{1}{2}}}
\def\sumi{\sum_{i=1}^{n}}
\def\summ{\sum_{i=2}^{n}}
\def\lst{\lambda_{\rm string}}
\def\umx{u_{\max}}

\def\npb{\nopagebreak}
\def\nn{\nonumber\\}
\def\sps{spherically symmetric\ }
\def\sss{static, spherically symmetric\ }

\newcommand{\vars}[1]{\left\{\begin{array}{ll}#1\end{array}\right.}

\bls{1.03}
\begin{document}
\begin{center}
               RUSSIAN GRAVITATIONAL SOCIETY\\
     CENTER FOR GRAVITATION AND FUNDAMENTAL METROLOGY, VNIIMS
\end{center}
\vskip 4ex
\begin{flushright}                 RGS-VNIIMS-001/95\\
                                   gr-qc/9505020
\end{flushright}
\vskip 15mm

\begin{center}
 {\bf ON SPHERICALLY SYMMETRIC SOLUTIONS\\[1ex]
 IN D-DIMENSIONAL DILATON GRAVITY}

\vskip2.5ex
     {\bf K.A.Bronnikov}\\
\vskip 5mm
     {\it Center for Gravitation and Fundamental Metrology, VNIIMS\\
     3--1 M. Ulyanovoy Str., Moscow 117313, Russia}\\
     e-mail: bron@cvsi.rc.ac.ru\\
\vskip 10mm
\end{center}

\noi{\bf ABSTRACT}

\bigskip\noi
     Exact static, spherically symmetric solutions to the
     Einstein-Abelian gauge-dilaton equations, in $D$-dimensional gravity
        with a chain of $n$ Ricci-flat internal spaces are considered, with the
        gauge field potential having three nonzero components: the temporal,
        Coulomb-like one, the one pointing to one of the extra dimensions, and
        the one responsible for a radial magnetic field. For dilaton coupling
        implied by string theory an $(n+5)$-parametric family of exact
        solutions is obtained, while for other dilaton couplings only
        $(n+3)$-parametric ones. The geometric
        properties and special cases of the solutions are discussed, in
        particular, those when there are horizons in the space-time. Two types
        of horizons are distinguished:  the conventional black-hole (BH) ones
        and those at which the physical section of the space-time changes its
        signature ({\it T-horizons}).  Two theorems are proved, one fixing the
        BH and T-horizon existence conditions, the other discarding the
        possibility of a regular center. Different conformal gauges are used to
        characterize the system from the $D$-dimensional and 4-dimensional
        viewpoints.
\vfill

\centerline{Moscow 1995}
\pagebreak
\twocolumn

\section{Introduction}
\markboth{1. Introduction}{1. Introduction}
        The search for and discussion of exact solutions to dilaton gravity
        equations has recently become the subject of many studies
        (\cite{dow,garf,gib,gibmaeda,jom,ksh,ksh2,sen} and many others)
        mostly because dilaton gravity (more precisely,
        Ein\-stein-ga\-uge-di\-la\-ton-axi\-on gravity) forms the bosonic part
of
        effective low energy string theory \cite{green}. Of the solutions
        in 4 dimensions existing to-date those found in Refs.\cite{myers94} and
        \cite{galz} are probably the most general: the former contains 6
        independent integration constants (the mass, dilaton, axion,
        electric and magnetic charges and the Taub-NUT parameter), the latter 5
        ones (the mass, electric and magnetic charges, the Taub-NUT and
rotation
        parameters; there are also asymptotic values of the dilaton and axion
        fields, which may be absorbed by proper re-definitions).

        This 4-dimensional approach assumes that the extra dimensions of the
        original (10-dimensional) theory are compactified and their papameters
        are constant. To take into account the possible
        variation of their size from point to point in the physical
4-dimensional
        space-time it is reasonable to adhere to multidimensional field
        equations, as is done, e.g., in \cite{br-ann,br-vuz91,br10,shira93}, or
        to include the extra-dimension scale factors as a separate dynamic
        variable \cite{cadmig}. Here the former approach is adopted. Moreover,
        we assume a sufficiently general space-time structure (see
        (\ref{Stru}) and an arbitrary value of the gauge-dilaton coupling
        constant $\lambda$ to cover a wider spectrum of possible
        multidimensional field theories, such as considered, e.g., in
        Ref.\cite{vladim}.

        Among the multidimensional solutions probably
        the most general is the solution of Ref.\cite{br10} containing $(n+4)$
        integration constants, where $n$ is the number of internal spaces (the
        mass, dilaton and electric charges and $(n+1)$ parameters connected
with
        the extra dimensions). Here the results of \cite{br10} are further
        generalized to include the magnetic charge; some properties of this
more
        general system are discussed.

        Unlike \cite{myers94,galz}, where symmetries of the field action were
        used to obtain the solutions from known, simpler ones, we try to
        integrate the field equations directly. One of the advantages of this
        approach is the possibility to consider more general field systems, for
        instance, with an arbitrary value of the metric-dilaton coupling
        constant $\lambda$.

        We will consider \sss
     Ein\-stein-Abe\-li\-an gauge-dilaton configurations in
$D$-di\-men\-si\-onal
     gravity with a chain of $n$ Ricci-flat internal spaces. We start from
     the action
\begin{equation}                                                     
     S= \int d^D x \sqrt{^D g}\Big[^D R+g^{MN}\varphi_{,M}\varphi_{,N}
               -\e^{2\lambda \varphi}F^2\Big]              \label{Action}
\end{equation}
{\sloppy
     where $g_{MN}$ is the $D$-dimensional
     metric,  $^D g=\bigl|\det g_{MN}\bigr|$,\ $\varphi$ is the dilaton scalar
     field and $F^2 = F^a F^a = F^{a\,MN}F^a_{MN},\
     F^a = dW^a,\ W^a\ (a= 1,\ 2, \ldots)$ being Abelian gauge fields of
     which one is to be interpreted as the electromagnetic field.

}
     Three types of $W^a$ compatible with spherical symmetry will be
     treated: $W^1$, the Coulomb-like one, so that the vector potential
     is $t$-directed, $W^2$, pointing to one of the extra dimensions, and
     $W^3$, responsible for a radial magnetic field.

     The field-theoretic limit of string
     theory corresponds to the specific value of the coupling constant
     $\lambda = \lst = \pm (D-2)^{-1/2}$ \cite{green,shira}.

     The field equations are written down in Sect.2 and solved in Sect.3.
     We come through a striking coincidence: if there is more than one
     nontrivial component of $F^a$, the field equations are explicitly
     integrable if and only if $\lambda^2 = 1/(D-2)$, i.e., exactly for the
     dilaton coupling which follows from string theory.

     In Sect.4 some special cases are indicated.
     In Section 5 we find special cases when the solutions exhibit
     black-hole or T-hole horizons and prove two theorems,
     valid for any values of $\lambda$.
     one determining the necessary conditions for horizon existence and
     the other on the nonexistence of solutions with a regular center in the
     model under study.

     Sects. 6 and 7 discuss different conformal gauges in
     $D$ and 4 dimensions, respectively. The point is that if the underlying
        theory is string theory, then a more fundamental role is played by the
        ``string metric'', or ``$\sigma$ model metric''
        $\g_{AB}=\e^{-2\lambda\varphi}g_{AB}$ rather than $g_{AB}$ from
        (\ref{Action}) (see, e.g., \cite{banks,shira} and references therein).
        Although mathematically a transition from $\g_{AB}$ to $g_{AB}$ may be
        treated as just a substitution simplifying the field equations,
        such issues as the nature of singularities (if any) and topology are
        better to discuss in terms of $\g_{AB}$. (Strictly speaking, this
        argument does not apply to $\lambda \ne \lambda_{\rm string}$ when the
        underlying more fundamental theory is not definitely fixed).

     On the other hand, the observable effects in 4 dimensions depend on how
     nongravitational matter interacts with the metric and dilaton fields and
     are described in different ways in different ``conformal gauges'', or
     systems of measurement, which are discussed in Sect.7. It should be
        stressed that such things as horizons and signatures are the same in
all
        the relevant conformal gauges since the conformal factors connecting
        them are regular at the horizons.

     Section 8 contains some concluding remarks.

     Throughout the paper capital Latin indices range form 0 to $D-1$, Greek
     ones from 0 to 3, the index $i$ enumerates subspaces and $a$ gauge field
     components.

\section{Field equations}
\markboth{2. Field equations}{2. Field equations}
     The set of field equations corresponding to (\ref{Action}) is
\begin{eqnarray}
 \nabla^M \nabla_M \varphi +\lambda \e^{2\lambda\varphi}F^2&=&0,
                                                            \label{Ephi}\\
         \ \ \nabla_N(\e^{2\lambda\varphi}F^{a\,NM})&=&0,\label{EMax}\\
     R_{MN}-g_{MN}R^A_A/2 &=& -T_{MN}                      \label{Einst}
\end{eqnarray}
     where $T_{MN}$ is the energy-momentum tensor
\begin{eqnarray}                                                       
&&\nq T_{MN} = \varphi_M\varphi_N - \half g_{MN}\varphi^A\varphi_A \nn
&&         + \e^{2\lambda\varphi}
     \big[-2F_M^{a\ A}F^a_{NA}+\half g_{MN}F^2\big].            \label{EMT}
\end{eqnarray}

     Consider a $D$-dimensional Riemannian or pse\-u\-do-Rie\-mann\-ian
manifold
     $V^D$ with the structure
\begin{eqnarray}                                                      
&& \nq    V^D = M^4 \times V_1\times \ldots \times V_n;  \nn
&&       \dim V_i=N_i; \quad D=4 + \sumi N_i,
                                                              \label{Stru}
\end{eqnarray}
     where $M^4 = M^2 \times S^2$
     is the conventional space-time and $V_i$
     are Ricci-flat manifolds of arbitrary dimensions and signatures
     with the line elements
     $ds_i^2$, $i=1,\ldots, n$. We seek \sss solutions to the field
     equations, so that the $D$-dimensional metric is
$$                                                            
       ds_D^2 = g_{MN}dx^M dx^N=     \qquad \qquad         $$
\begin{equation}
 \e^{2\gamma(u)}dt^2-\e^{2\alpha(u)}du^2 - \e^{2\beta(u)}d\Omega^2
          + \sumi \e^{2\beta_i(u)}ds_i^2.                       \label{DsD}
\end{equation}
     where $d\Omega^2=d\theta^2 + \sin^2\theta d\phi^2$ is the line element
     on a unit sphere $S^2$, while all the scale factors $\e^{\beta_i}$ of
     the internal spaces $V_i$ depend on the radial coordinates $u$.

     It should be noted that in (\ref{DsD}) one could include arbitrary
     $(d+1)$-dimensional spheres; however, a nonzero magnetic field $W^3$ is
     compatible (at least in the conventional approach) only with $d=1$.
     Solutions with $W^3=0$ and arbitrary $d$ have been considered in
     \cite{br10}.

     If we denote $\gamma =\beta_{-1},\ N_{-1}=1,\ \beta=\beta_0,\ N_0=2$
     and choose the harmonic radial coordinate $u$ \cite{br-acta} such that
\begin{equation}
     \alpha = \sum_{i=-1}^{n}N_i \beta_i \equiv \gamma+2\beta +\sigma,
              \quad  \sigma\equiv\sumi N_i\beta_i,         \label{Harm}
\end{equation}
     the Ricci tensor components $R_M^N$ can be written in the highly
     symmetric form ($x^1\equiv u$)
\begin{eqnarray}                                                       
     R_u^u \al=\al -\e^{-2\alpha}\sum_{i=-1}^{n}N_i [\beta_i''
       +\beta_i^{'2}-\beta_i'\alpha']; \nn
     R_{\mu}^M \al=\al 0 \qquad (M>d+2;\ \ \mu =0,\ldots ,d+2); \nn
     R_{a_j}^{b_i} \al=\al
     -\delta_j^i\delta_{a_i}^{b_i}\e^{-2\alpha}\beta_i'', \qquad
                    i\ne 0                                  \nn
   R_\theta^\theta \al=\al R_\phi^\phi = 2\e^{-2\beta}-\e^{-2\alpha}\beta'';
                                                              \label{Ricci}
\end{eqnarray}
     where primes denote $d/du$ and the indices
     $a_j\ (b_i)$ refer to the subspace $V_j\ (V_i)$.

     The fields $\varphi$ and $F$ compatible with the assumed symmetry
     are $\varphi=\varphi (u)$, a component of $W^a$ in the $t$
     direction (the Coulomb electric field, $W^1 = w^1(u)dt$), a similar
     components in some internal one-dimensional subspace (if any;
     this subspace will be denoted $V_1$ and parametrized by a coordinate
     $v$: $N_1=1,\ \beta_1=\nu (u),\ W^2 = w^2(u)dv$)  and
     a monopole magnetic field of the form $W^3 = \qq \cos\theta\, d\phi$
     where $\qq$ is the magnetic charge.

     The gauge field strengths are
\begin{eqnarray*}
 \al\al\nqq  F^a = F^a_{AB}dx^A\wedge dx^B:
    \qquad F^1 = {w^1}'(u)\,du\wedge dt;\\
    \al\al F^2 = {w^2}'(u)\,du\wedge dv;
    \qquad F^3 = -\qq \sin \theta \, d\theta\wedge d\phi.
\end{eqnarray*}
     The vector field equations lead to
\begin{eqnarray}                                                      
     \e^{2\lambda\varphi+2\alpha}F^{1\,ut} = q = \const,\nn
     \e^{2\lambda\varphi+2\alpha}F^{2\,uv} = q'= \const.         \label{Maxw}
\end{eqnarray}
     Here $q$ is the electric charge and $q'$ is the
     gauge charge connected with $F^2_{uv}$. We will sometimes also use the
        notations $q=q_1,\ q'=q_2,\ \qq=q_3$.

     Now the scalar field equation and some linear combinations of the metric
     field equations may be written in the form
\begin{eqnarray}                                                   
\nqq    \half (N+1)\gamma''\al=\al
     N q^2\e^{2\omega} -\eta_v{q'}^2\e^{2\psi} +\qq^2\e^{2\chi},
\label{Egamma}\\[3pt]
\nqq     \half (N+1)\nu'' \al=\al
     - q^2\e^{2\omega} +N\eta_v{q'}^2\e^{2\psi}+\qq^2\e^{2\chi},
                                                            \label{Enu}\\[3pt]
\nqq    \half  (N+1)\beta_i''\al=\al
     - q^2\e^{2\omega} -\eta_v{q'}^2\e^{2\psi}+\qq^2\e^{2\chi},\nn
                &&\qquad  i=2,\ldots, n;                \label{Ebetai} \\[3pt]
\nqq     \varphi''\al=\al (N+1)\lambda\beta_i'',           \label{Ephi1}\\[5pt]
\nqq     (\alpha-\beta)''\al=\al \e^{2\alpha-2\beta},          \label{Ebeta}
\end{eqnarray}        \vspace{-7mm}
\begin{eqnarray}
  {\alpha'}^2-\sum_{i=-1}^{n}N_i{\beta_i'}^2
       -2\e^{2\alpha-2\beta}\hspace{15mm}\nn
     = \varphi'^2 - 2q^2\e^{2\omega}-2\eta_v {q'}^2\e^{2\psi}
                -2\qq^2\e^{2\chi}                               \label{Int}
\end{eqnarray}
     where we have denoted
\begin{eqnarray}                                                       
&& \nqq    \omega= \gamma-\lambda\varphi;\quad
         \psi= \nu-\lambda\varphi;\quad
        \chi = \gamma+\sigma+\lambda\varphi; \nn
&&  \nq  N = D-3;  \qquad      \eta_v = \sign g_{vv}.\quad     \label{Defopsi}
\end{eqnarray}
     Eq.(\ref{Int}) is the $(^u_u)$ component of the Einstein equations and
     represents a first integral of (11)-(15).
\medskip

\section{Exact static solutions}
\markboth{3. Exact static solutions}{3. Exact static solutions}
     Eq.(\ref{Ebetai})  is easily integrated to give
\begin{equation}
    \e^{\beta-\alpha} = s(k,u) \equiv
     \vars{ k^{-1}\sinh\,ku,  & k>0; \\
     u, &  k=0;  \\
     k^{-1}\sin ku,  & k<0 }                           \label{Beta}
\end{equation}
     where $k=\const$ and an inessential integration constant has been
        eliminated by shifting the origin of $u$. Consequently,
        with no loss of generality one can assert that the harmonic
        $u$ coordinate is defined for $u>0$ and $u=0$ corresponds to spatial
        infinity. By (\ref{Beta}), asymptotically ($u\to 0$) the
        conventional flat-space spherical radial coordinate $r=\e^{\beta}$ is
        connected with $u$ by $u = 1/r$.

     The remaining equations may be combined into a set written in terms of
     the functions $\omega,\ \psi,\ \chi$ in the following matrix form:
\bear
\al\al    \frac{N+1}{2}
          \left[ \begin{array}{c}
          \omega'' \\ \psi'' \\ \chi'' \end{array} \right] \nn
\al\al\nqq {} =  \left[ \begin{array}{ccc}
             N+\Lambda & \Lambda -1 & 1-\Lambda \\
             \Lambda -1 & N+\Lambda & 1-\Lambda \\
             1-\Lambda &  1-\Lambda & N+\Lambda \end{array}\right]
     \left[ \begin{array}{r}
     q^2\e^{2\omega}\\ \eta_v {q'}^2\e^{2\psi}\\ \qq^2\e^{2\chi}
              \end{array} \right]                           \label{Ematr}
\ear
     where $\Lambda = \lambda^2(N+1) =(\lambda/\lst)^2$.
     The matrix in the right-hand side is nondegenerate for all $\Lambda$ if
     $N\geq 2$; at $N=1$ there is no variable $\nu$ (and $\psi\equiv
     \nu-\lambda\varphi$), nor charge $q'$, so that the second line in Eq.%
     (\ref{Ematr}) and the second column in the matrix must be removed.

     Eqs.(\ref{Ematr}) can be explicitly integrated either if there is only
     one nonzero charge among $q,\ q',\ \qq$, or in the case $\Lambda=1$,
     exactly the one corresponding to the dilaton coupling constant in string
     theory. The functions $\gamma,\ \nu,\ \varphi,\ \beta_i$ are then easily
     expressed in terms of $\omega,\ \psi,\ \chi$. We will here present all
     these solutions for information purposes, although Solutions E
     (``electric''), I (``internal'') and G (``general'', for $\Lambda=1$)
     have been given earlier (see \cite{br10} and references therein), the
     latter only for $\qq=0$, in slightly different notations. The fields
     $F^a$ are in all cases expressed by (\ref{Maxw}).
     In addition, inessential integration constants are eliminated by
     rescaling the coordinates in the subspaces $V_i$: their true scales are
     thus hidden in $ds_i^2$, while the factors $\e^{\beta_i}$ are normalized
     by $\beta_i (0)=0$. We impose one more condition: $\varphi(0)=0$;
     this is also no generality loss since a nonzero $\varphi(0)$ can be
     compensated by rescaling the charges $q_a$.

     The number of essential integration constants is in all
     cases equal to $n$, the number of internal spaces, plus the number of
        nontrivial physical fields:  scalar, gauge ($F^a$) and gravitational.
        The constants are denoted by, respectively, $b_i,\ C,\ q_a,\ k,\ h_a$.
\medskip

     {\bf Solution E}\ \cite{potsdam,bm-itogi,br10,br-vuz91} \npb
\nx
\begin{eqnarray}                                                    
\nqq     ds_D^2 \al = \al \e^{2\gamma}dt^2 -
 \e^{-2\gamma/N}\Biggl\{\left[\frac{\e^{-Bu}}{d\cdot s(k,u)}\right]^{2/d}\nn
\nqq \al \times \al \left[\frac{du^2}{d^2 s^2 (k,u)}+d\Omega_{d+1}^2\right]
     -\sumi \e^{2b_i u}ds_i^2\Biggr\},                   \label{DsE}\\
\nqq     \varphi \al=\al Cu/A -2\lambda N_{+}\omega,
\label{PhiE}\\
\nqq     \gamma \al=\al (\omega +\lambda Cu)/A,
\label{GammaE}\\
\nqq     \e^{-\omega}\al=\al Q_1 s(h_1,u+u_1);
     \ u_1={\rm const};\ \omega(0)=0.                    \label{OmE}
\end{eqnarray}
     The integration constants are connected by the relation due to
     (\ref{Int})
\begin{equation}                                                    
     2k^2\sign k = 2N_{+}h_1^2 \sign h_1 +
     \frac{C^2}{A}+B^2+\sumi N_i b_i^2.                \label{IntE}
\end{equation}          \nx
     The notations are                                              
\bear
     Q_1=\sqrt{q^2/N_+},\qquad A = 1+\Lambda/N, \nn
     N_+=(N+1)/(N+\Lambda), \qquad
         B=\sumi N_i b_i    \                            \label{Defconst}
\ear
     The last condition from (\ref{OmE}) is the requirement that $\gamma
     =0$ at infinity, i.e., $dt$ is a time interval measured by a distant
     observer at rest with respect to our static configuration.  One may
     notice that no function $\nu$ is distinguished among the
     scale factors $\beta_i$.

\medskip
     {\bf Solution I}\ \ \cite{br10}   \npb
\medskip

     The solution can be found by proper substitutions in Solution E.
     Namely, we obtain:
\begin{eqnarray}                                                
\nqq     ds_D^2\al = \al \eta_v\e^{2\nu}dv^2
           + \e^{2\nu/N}\biggl\{\e^{2b_0 u}dt^2 \nn
\nqq     \al\al\nqq {} -\frac{\e^{-2Bu}}{s^2(k,u)}
     \biggl[\frac{du^2}{s^2 (k,u)}+d\Omega^2\biggr]
     +\summ \e^{2b_i u}ds_i^2\biggr\},                   \label{DsI} \\
\nqq     \varphi \al=\al Cu/A -2\lambda N_{+}\psi,               \label{PhiI}\\
     \nu \al=\al (\psi +\lambda Cu)/A,                       \label{NuI}\\
     \e^{-\psi} \al = \al \vars{
          Q_2  s(h_2, u+u_2)\ \ &{\rm if}\ \ \eta_v=+1,\\
          (Q_2/h_2)\cosh h_2(u+u_2) &{\rm if}\ \ \eta_v=-1,}\nn
\nq  \al\al\cm         \psi(0)=0                                  \label{PsiI}
\end{eqnarray}
     where $Q_2= \sqrt{q'^2/N_+}$.
        The integration constants are constrained by
\bear                                                                
 \nqq    2k^2\sign k \al = \al 2N_{+}h_2^2 \sign h_2 \nn
 \nqq   \al \al {} + \frac{C^2}{A}+B^2+b_0^2+\summ N_i b_i^2.      \label{IntI}
\ear
     In (\ref{DsI}) and (\ref{IntI}) $B$ is expressed as
     $B=b_0+\sum\limits_{i=1}^n N_ib_i$.

\medskip
     {\bf Solution M}                  \npb   \nx
\bear                                                          
 \nqq    ds_D^2 \al =\al \e^{2\gamma}dt^2 -
     \frac{\e^{-2N\gamma-2bu}}{s^2(k,u)}
     \left[\frac{du^2}{s^2 (k,u)}+d\Omega^2\right] \nn
   && \cm {+} \e^{2\chi/(AN)}\sumi \e^{2b_i u}ds_i^2;            \label{DsM}\\
   \varphi \al=\al Cu + 2\lambda N_+ \chi;                   \label{PhiM}\\
   \gamma  \al=\al b_0 u + \chi/(AN);                       \label{GammaM}\\
   \e^{-\chi} \al=\al \qq^2 s(h_3, u+u_3),\ \ \ \chi(0)=0,   \label{ChiM}
\ear
\bear  \nqq 2k^2\sign k \al = \al 2N_+ h_3^2\sign h_3 \nn
       \nqq    \al + \al C^2 (1+\lambda^2) +b_0^2 + \sumi N_i b_i^2.
\label{IntM}
\ear
     where $b_0 = -\lambda C - b,\ b=\sumi N_i b_i, Q_3=\sqrt{\qq^2/N_+}$.
\medskip

     {\bf Solution G} ($\lambda^2 = \lst^2 = N+1$)       \npb  \nx
\begin{eqnarray}                                                   
\nqq \nqq    ds_D^2 \al=\al \e^{2\lambda\varphi}\biggl\{
     \e^{2\omega}dt^2 -
     \frac{\e^{-2\chi}}{s^2(k,u)}
       \biggl[\frac{du^2}{s^2(k,u)}+d\Omega^2 \biggr]  \nn
\nqq \al\al\cm {} +\eta_v \e^{2\psi}dv^2 + \summ \e^{2b_i u}ds_i^2\biggr\}
                                                            \label{DsG}\\
     \e^{-\omega}\al=\al Q_1 s(h_1, u+u_1),\ \ \ \omega(0)=0;   \label{OG}\\
     \e^{-\psi} \al=\al \vars{
                 Q_2 s(h_2, u+u_2),\ \al \eta_v=+1,\\
             (Q_2/h_2)\cosh h_2(u+u_2),\ \al  \eta_v=-1;}       \label{PsiG}\\
     \e^{-\chi}\al = \al Q_3 s(h_3, u+u_3),\ \chi(0)=0;  \label{ChiG}\\
     \lambda\varphi \al = \al (\chi-\omega-\psi-bu)/(N+1);
                \label{PhiG}\\
 \al\al\nqq  2k^2\sign k \sum_{a=1}^{3}h_a^2\sign h_a +\summ N_i b_i^2.
                                                         \label{IntG}
\end{eqnarray}
     where $Q_a=\sqrt{2q_a^2},\ a=1,2,3$, $\psi(0)=0$ and $b=\summ N_i b_i$.

     The ``intermediate'' solutions for $\Lambda=1$ and two nonzero charges
     (to be labelled EM, EI, IM) are easily obtained from Solution G in the
     corresponding limits. The only subtle point is that the limit, say,
     $Q_1\to 0$ can be realized only if $h_1>0$ and,
     moreover, the constant $u_1$ must vary along with $q_1\to 0$ so as to
     maintain $\omega(0)=0$. Thus in the limit $Q_1=0$ we obtain
     $\omega=-h_1 u$, just what could be obtained directly from the equations
     with $q=0$. The situation is the same with all limits $q_a\to 0$, for
     instance, when obtaining the purely scalar-vacuum solution from either
     of the solutions E, I, M. In particular, the EI solution obtained from
     Solution G in this way coincides with Solution C from Ref. \cite{br10} up
     to notations of some constants.

\section{Special cases}
\markboth{4. Special cases}{4. Special cases}
     Let us enumerate some special cases of the solutions.      \npb
\begin{description}
\item[(a)]
     Either of the solutions yields the well-known scalar-vacuum solution in
     $D$ dimensions if the gauge fields $F^a$ are switched off \cite{briv,fim}.
     Their specialization to 4 dimensions leads to the scalar-vacuum
\cite{fish}
     and vacuum (Schwarzschild) solutions of general relativity.
\item[(b)]
     When $\lambda =0$, Solution E reduces to the generalized
     Reissner-Nordstrom (RN) one for linear scalar and
     electromagnetic fields \cite{br-ann};
     at $D-4,\ n=0$ it coincides with the Penney
     solution \cite{penney} which in turn reduces to the RN
     one when the scalar field is eliminated.
\end{description}
     A new feature implied by nonzero gauge fields as compared with item
     (a) is that the constants $k$ and (or) $h_a$ can have either sign and
     the functions $s(h_a, u+u_a)$ can be sinusoidal, leading to
     $u_{\max}<\infty$. In Solution E that means the
     appearance of a RN-like repelling singularity at the center of the
     configuration.
\begin{description}
\item[(c)]
     Solution E, if deprived of extra dimensions, yields the solution
     for interacting scalar and electromagnetic fields in
     general relativity, first obtained in Ref.\cite{brsh-vuz}.
\item[(d)]
     The case $D=5\ (n=N_1=1)$ with no dilaton and gauge fields was
     considered in Refs.\cite{kramer,davow} and many others; it
     also coincides with the ``Kaluza-Klein soliton'' considered by Gross and
     Perry \cite{groper}. Similar solutions for $D=6$ and $D=7$ are presented
     in \cite{vladim}; see also references therein.
\item[(e)]
     There are special cases of the above solutions when the space-time
     exhibits horizons. They are discussed in the next section.
\end{description}
\section{Horizons: black holes and time holes}
\markboth{5. Horizons: black holes and time holes}
         {5. Horizons: black holes and time holes}
\subsection{Horizons in Solution E}
     The behavior of the metric
     for different combinations of integration constants is rather various.
     However, calculations show that, in particular, Solution E has a naked
     singularity at $u=\umx$ or $u=\infty$ in all cases except
\begin{equation}
  b_i=-\frac{k}{N}; \quad h_1=k>0;
      \quad C=-\lambda k\frac{N+1}{N},\label{ConBH}
\end{equation}
     when the sphere $u=\infty$ is a Schwarzschild-like event horizon:
     at finite radius $r=\e^{\beta}$ of a coordinate sphere
     the metric coefficient $g_{tt}=0$ and the light travel
     time $\int \e^{\alpha -\gamma} du$ diverges \cite{br-vuz91}.

     In this black-hole (BH) case only two independent
     integration constants remain, say,  $k$ and $q$, and the coordinate
     transformation
\begin{equation}
     \e^{-2ku} = 1-2k/R                                     \label{TransR}
\end{equation}
     brings the solution to the form \cite{br-vuz91}
\begin{eqnarray}                                                    
     ds_D^2 \al=\al \frac{(1-2k/R)dt^2}{(1+p/R)^{2/A}}         \nn
     -\al\al\nq (1+p/R)^{2/AN}\left[\frac{dR^2}{1-2k/R}
                 +R^2 d\Omega^2 - \sumi ds_i^2\right];                \nn
     \e^{\lambda\varphi} \al=\al (1+p/R)^{\Lambda/(\Lambda+N)};\nn
     F \al= \al F^1 = q(R+p)^{-2} dR\wedge dt;                        \nn
     p \al=\al (k^2+q^2/N_+)^{1/2} - k.                          \label{BHE}
\end{eqnarray}
     This extends the well-known dilaton
     black-hole solution (see, e.g., \cite{gibmaeda,garf,hein}) to spaces of
     the form (\ref{Stru}). For the first time it was obtained in 4
        dimensions \cite{brsh-vuz} where for $\lambda=0$ it reduces to the RN
        one. More frequently used notations are connected with ours by
\begin{equation}                                                      
     R+p = r;\qquad  p=r_-;\qquad 2k= r_+ - r_-.
        \label{r+}
 \end{equation}

     In this family of BH solutions a nonzero dilaton
     field exists solely due to the interaction ($\lambda\ne
     0$).  When $\lambda=0$, i.e., the $\varphi$ field becomes minimally
     coupled, a horizon is compatible only with $\varphi=$const. This
     conforms with the well-known ``no-hair'' theorems and the properties of
     the general-relativistic scalar-vacuum and scalar-electrovacuum
     configurations.
\subsection{Horizons in Solution I}\npb
     There are counterparts of the above BH solutions among those
     of Class I. Namely, under the same conditions (\ref{ConBH}) (where just
     $h_1$ is replaced by $h_2$) the sphere $u=\infty$ is a horizon as well
     but now in the $(v,u)$ subspace instead of $(t,u)$ in the previous,
     conventional case. The solution is
\begin{eqnarray}                                                      
     ds_D^2 \al = \al \frac{(1-2k/R)\eta_v dv^2}{(1+p'/R)^{2/A}}
          + (1+p'/R)^{2/AN}                                   \nn
        \al \times \al \bigg[dt^2- \frac{dR^2}{1-2k/R}
          - R^2 d\Omega_2^2 + \summ ds_i^2\bigg],              \nn
     \e^{\lambda\varphi} \al=\al (1+p'/R)^{\Lambda/(\Lambda+N)};\nn
     F \al=\al  q'(R+p')^{-2} dR\wedge dv,                       \nn
     p' \al=\al (k^2+\eta_v q'^2/N_+)^{1/2} - k.               \label{TH}
\end{eqnarray}

     The main feature of these configurations is that the physical space-time
     $M^4$ changes its signature at $R=2k$: it is $(+---)$ at $R>2k$ and
     $(++--)$ at $R<2k$. This evidently means that the anomalous domains
     should contain quite unconventional physics whose possible
     consequences and observational manifestations are yet to be studied. It
     has been suggested \cite{br10} to call the domains with an unusual
     space-time signature {\sl time holes} or {\sl T-holes} and the
     corresponding horizons {\sl T-horizons}.

     Each BH configuration of any dimension $D>4$ has
        a family of T-hole counterparts (a family since the subspaces
     $V_i$ may have different dimensions and signatures) and {\sl vice
     versa}. However, if a BH possesses an external field, such as
     the Coulomb field of a RN black hole, its T-hole analog has a field
     modified  by the $t\chg v$ interchange, as is the case with
     the above solutions: the Coulomb-like field becomes the one pointing
        in the $v$ direction which from the 4-dimensional viewpoint looks like
a
     scalar field interacting with the dilaton (see Sect.7).

     Unlike a BH-horizon, a T-horizon is not in absolute past or
     future from a distant observer's viewpoint, it is visible since it takes
     a finite time for a light signal to come from it
     ($\int\e^{\alpha-\gamma}du < \infty$, independently of a conformal
     gauge).    T-hole properties are further discussed in Sect.6.



\subsection{Horizons in Solutions M and EM}
            Solution M admits a BH horizon if and only if
\begin{equation}                                                    
     h_3=k>0,\ \, b_i = k/(N+\Lambda),\ \,
          \lambda C=\Lambda k/(N+\Lambda)             \label{ConBH-M}
\end{equation}
     After the same substitution (\ref{TransR}) the solution takes the form
\begin{eqnarray}
     ds_D^2 \al=\al (1+\pp/R)^{-2/AN} \sumi ds_i^2 +
               \frac{(1-2k/R)dt^2}{(1+\pp/R)^{2/AN}}\nn            
             \al - \al (1+\pp/R)^{2/A}\biggl[\frac{dR^2}{1-2k/R}
               +R^2 d\Omega^2\biggr];                         \nn
     \e^{\lambda\varphi} \al=\al (1+\pp/R)^{-\Lambda/(\Lambda+N)};\nn
     F \al = \al F^3 = -\qq \sin\theta\, d\theta\wedge d\phi;
\nn
     \pp \al=\al (k^2+\qq^2/N_+)^{1/2} - k.                       \label{BHM}
\end{eqnarray}
     Like (\ref{BHE}), this solution is well-known for $D=4\ (N=1,\ n=0)$. It
     is easy to notice that just for $N=1$ the metrics (\ref{BHE}) and
     (\ref{BHM}) coincide, while the $\varphi$ fields are different. At
     higher dimensions the distinctions are more complicated.

     The same solutions turns into a T-hole one if one replaces, as before,
     $t\chg v$. This case is simpler than that of Solutions E and
     I since the interchange does not affect the magnetic field.

     Solution EM (with $\lambda=\lst$, obtained from Solution G in the limit
     $q'\to 0$) also has a BH  horizon under the conditions
\begin{equation}                                                       
     h_1=h_3=k>0;\ \  b_i=0\ \ (i=1,\ldots,n).          \label{ConDy}
\end{equation}
     The substitution (\ref{TransR}) leads to the following form of the
     solution:
\begin{eqnarray}                                                       
     ds_D^2 \al=\al \e^{2\lambda\varphi}\biggl\{
     \frac{1-2k/R}{(1+p/R)^2}dt^2                             \nn
     \al - \al (1+\pp/R)^2\biggl[\frac{dR^2}{1-2k/R}
               +R^2 d\Omega^2\biggr] + \sumi ds_i^2\biggr\};      \nn
     \e^{2\lambda\varphi}\al=\al
       \biggl(\frac{1+p/R}{1+\pp/R}\biggr)^{2/(N+1)};\nn
     F \al = \al q(R+p)^{-2}dR\wedge dt - \qq \sin\theta \, d\theta\wedge
d\phi;\nn
  \al\al \nqq\nq p=(k^2+2q^2)^{1/2} -k;\quad
     \pp=(k^2+2\qq^2)^{1/2} -k.                         \label{Dyon}
\end{eqnarray}
     This is a dyon BH solution in dilaton gravity of arbitrary dimension,
     which naturally passes to (\ref{BHE}) and (\ref{BHM}) with
     $\lambda=\lst$ in the limits $\qq\to 0$ and $q\to 0$, respectively.

     To obtain the T-hole counterpart of (\ref{Dyon}) (a special case of
     Solution IM) one has to change $t\chg v,\ q\to q',\ p\to p'$.
\subsection{The general case. Non-existence theorems}
    It can be directly verified that among the Class G solutions
    with three nonzero charges $q_a$
    there are no special cases with horizons, either
    BH- or T-hole ones. This observation can be generalized to include
    the cases $\lambda\ne\lst$ when exact solutions are hard to obtain.
    Namely, we will prove a theorem generalizing Theorem 1 of Ref.
    \cite{br10} (the latter concerned configurations with nonzero $q$ and
    $q'$ but $\qq=0$). The present proof, employing directly the horizon
    regularity condition, is much simpler than that in \cite{br10}.

    Let us previously adopt a convenient horizon definition for our static,
    spherically symmetric configurations. Namely, we will call a {\sl
    BH-horizon} (i.e., a conventional black-hole horizon) a
    nonsingular sphere in
    a space with the metric (\ref{DsD}) where the metric
    functions $\beta$ and $\beta_i$ are finite while
    $g_{tt}= \e^{2\gamma} \to 0$. Similarly, a {\sl T-horizon} is a
    nonsingular sphere
    where $\beta,\ \gamma,\ \beta_i\ (i=2,\ldots,n)$ are finite
    while $g_{vv}=\e^{2\nu}\to 0$ where $v$ parametrizes the one dimensional
    subspace $V_1$, one of the internal subspaces.
    (We ought to require that, in addition, the
    travel time $\int \e^{\alpha-\gamma}du$ of a light signal approaching
    it be infinite; however, this condition is irrelevant for our
    argument.)

\medskip
{\bf Theorem 1.}
{\sl The static, spherically symmetric field system}
     (\ref{Action},\ref{Stru},\ref{DsD},\ref{Maxw}) {\sl has no BH horizon if
     $q'\ne 0$ and has no T-horizon if $q\ne 0$.}
\medskip

\noindent
{\it Proof.}\ Assume that $q'\ne 0$ and there is a BH horizon at some
     $u=u^*$. By (\ref{Beta}) and (\ref{Harm}), $\beta=-\gamma-\sigma
        -\ln s(k,u)$, whereas $\beta$ and $\sigma$ must be finite at $u=u^*$
and
        $\gamma\to\infty$. This implies
\begin{equation}                                                       
        \ln s(k,u) = -\gamma(u) + O(1),                         \label{Lns}
\end{equation}
        whence $s\to\infty$ at $u\to u^*$. By definition of $s(k,u)$, this is
        possible only if $k\geq 0,\ u^*=\infty$.

        The metric regularity condition at $u\to\infty$ implies, in particular,
        that the invariant $|R^M_N R^N_M| <\infty$. In our case the Ricci
tensor
        $R^M_N$ is diagonal and connected with the energy-momentum tensor by
        (\ref{Einst}). Consequently, the above invariant is just a sum of
        squares and its finiteness implies the finiteness of each
        summand. As the invariant $R_M^M$ is also finite, each component of
        $T^M_N$ is finite as well. Recalling the explicit expressions for
        $T^M_N$, we obtain the requirements:
\begin{eqnarray}                                                      
{\rm (a)}\ \e^{-2\alpha}{\varphi'}^2 <\infty;\quad
{\rm (b)}\ q^2\e^{-2\alpha+2\omega} <\infty;\nn
{\rm (c)}\ {q'}^2\e^{-2\alpha+2\psi} <\infty;\quad
{\rm (d)}\ \qq^2\e^{-2\alpha+2\chi} <\infty.                    \label{QQ}
\end{eqnarray}
        If $k>0$, from (\ref{Lns}) it follows $\gamma=-ku+O(1)$ at $u\to\infty$
        and $\alpha(u)$ has the same asymptotic. Then by (\ref{QQ}.a)
$\varphi'$
        decays exponentially and $|\varphi(\infty)|<\infty$, so that $|\psi| =
        |\nu-2\lambda\varphi| <\infty$. As $\alpha\to\infty$, we can satisfy
        (\ref{QQ}.c) only if $q'=0$, in contrast to what was assumed.

        The case $k=0$ is somewhat more involved. At $u\to\infty$ by
(\ref{Lns})
        $\gamma=-\ln u +O(1)$ and the same asymptotic has $\alpha(u)$. By
        (\ref{QQ}.a) then $\varphi'=O(1/u)$ and $\varphi= C\ln u +O(1)$ where
        $C$ is a constant. The finiteness condition (\ref{QQ}.c) only implies
        then that either $q'=0$, or $\lambda C \geq 1$.

        In the latter case (\ref{QQ}.d) can be satisfied only if
        $\qq=0$. Therefore let us assume that $q'\ne 0,\ \qq=0,\ \lambda C\geq
        1$ and address to Eqs.(\ref{Egamma}) and (\ref{Enu}) at the asymptotic
        $u\to\infty$. As $\e^{\omega}\sim u^{-1-\lambda C}$ and $\e^{\psi}\sim
        u^{-\lambda C}$, in both equations the leading terms are those with
        ${q'}^2$ and, subtracting them, one can write: $N\gamma''
        -\nu''=O(u^{-2-2\lambda C})$. Since $2+2\lambda C \geq 4$, we conclude
        that $N\gamma-\nu$ at the asymptotic $u\to\infty$ can be either a
        constant, or a linear function of $u$, both possibilities being
        inconsistent with $|\nu(\infty)|<\infty$ and
        $\gamma=-\ln u +O(1)$.

     There is still one more possibility, namely, that
     the map with the coordinate $u$ is
     incomplete in the present static frame of reference. This may happen if
     $u=\infty$ is a regular surface and another coordinate must be used to
     penetrate beyond it, where anything may be found, in particular, a
        horizon.  This is, however, not the case for our system.
        Indeed, assume that $\beta,\ \beta_i,\ \nu$ and $\gamma$ are finite at
        $u=\infty$. Then from (\ref{Beta}) it
     follows that $s(k,u)$ must have a finite limit at $u\to\infty$,
     contrary to the definition of $s(k,u)$.

     Thus a BH-horizon is inconsistent with $q'\ne 0$.
        By symmetry of our equations with respect to
        $\gamma$ and $\nu$, a T-horizon is inconsistent with $q\ne 0$.  (The
        only asymmetry, the possibility of $\eta_v=-1$, is insignificant for
the
        above argument).  The theorem is proved. $\bullet$

     Another general statement valid for all $\lambda$ is

\medskip
{\bf Theorem 2.} {\sl The field system}
     (\ref{Action},\ref{Stru},\ref{DsD},\ref{Maxw}) {\sl cannot form a
     \sss configuration with a regular center.}
\medskip

     A {\sl proof} makes use of the same type of
     argument as that of Theorem 1. $\bullet$
\medskip

{\sl Remark.} A regular center assumes the regularity
     of $g_{MN}$ and $\varphi$. Consequently,
     the statement of Theorem 2 is valid as well for all other
     conformal gauges connected with $g_{MN}$ by factors of the form
     $\exp(\const\cdot\varphi)$. The same is true for Theorem 1 if the
     (natural) additional requirement is adopted that $\varphi$ should be
     finite at a horizon.

     Thus, although our system consists of interacting fields, they cannot
     create a particle model with a regular center. On the other hand,
     regular spherical configuration with no center at all,
     like wormholes or ``cornucopions'' \cite{banks}
     are certainly not excluded, as seen, e.g., from the next section.

\section{String metric}                                       
\markboth{6. String metric}{6. String metric}
\subsection{Properties of Solution G}
     As already pointed out, in the case
     $\lambda=\lst= \pm\sqrt{N+1}$ it is more adequate to study the field
     behavior in terms of the so-called string metric
\begin{equation}                                                
        \g_{MN}=\e^{-2\lambda\varphi}g_{MN}
\label{MapS}
\end{equation}
     rather than the metric $g_{MN}$ in the ``Einstein gauge''
     (such that the coefficient by $^D R$ in the Lagrangian is constant),
     the most convenient one for solving the equations.
     For other $\lambda$ the same can be done by analogy, leading to the
     action (\ref{Action}) in terms of $\g_{MN}$
\begin{eqnarray}                                                       
\nqq\nqq\al\al S= \int d^D x \sqrt{\g^D}\e^{(N+1)\lambda
\varphi}\Big\{\hat{R}^D
                                                     \nn
\nqq\nqq  \al\al\ {} + [1-\lambda^2(N+1)(N+2)]\g^{MN}\varphi_{,M}\varphi_{,N}
               -\hat{F}^2\Big\}                              \label{ActS}
\end{eqnarray}
     where symbols with hats denote quantities obtained with or corresponding
     to $\g_{MN}$.

     Nevertheless, we will discuss some features of the solutions
     only in the case $\lambda=\lst$ for which the metric $\g_{MN}$ is
     manifestly meaningful. Moreover, as we saw in Sect.3, $\lambda=\lst$
     is the condition under which solutions with more than component of $F^a$
     can be obtained.

     For $\lambda=\lst$ we can address to the general Solution G, all
     the others being its special cases. It is described by Eqs.%
     (\ref{OG})-(\ref{IntG}) with the string metric given by the expression
     in curly brackets in (\ref{DsG}).

     For $\eta_v=1$ Solution G may be written as follows:
\begin{eqnarray}                                                
 \al\al       \ds = \frac{dt^2}{2q^2 s^2(h_1,u+u_1)}             \nn
 \al\al   {} -\frac{2\qq^2 s^2(h_3,u+u_3}{s^2(k,u)}
               \biggl[\frac{du^2}{s^2(k,u)}+ d\Omega^2\biggr] \nn
 \al\al   {} +\frac{\eta_v dv^2}{2{q'}^2 s^2(h_2,u+u_2)}
           +\summ \e^{2b_i u}ds_i^2;                          \label{DsGS}
\end{eqnarray} \nxx\nx
\begin{equation}                                                 
   2k^2\sign k = \sum_{a=1}^{3}h_a^2\sign h_a +\summ N_i b_i^2;
                                                           \label{IntGS}
\end{equation}
     the fields $\varphi$ and $F^a$ are determined by (\ref{PhiG}) and
     (\ref{Maxw}), respectively.

     The coordinate $u$ is defined in the range
\begin{equation}                                                   
     0 < u \leq\umx = \min\bigl\{z(s_0), z(s_1), z(s_2),z(s_3),\infty\bigr\}
                                                                 \label{Umax}
\end{equation}
     where $u=0$ corresponds to spatial infinity, $s_0=s(k,u),\ s_a=
     s(h_a,u+u_a)\ (a=1,2,3)$ and $z(s)$ is the smallest positive zero of the
     function $s$.

     Thus the solution behaviors for different sets of integration constants
     are naturally classified by the variants of $\umx$ by (\ref{Umax}).
     We will label them by boldface figures from {\bf 0} to {\bf 4},
        respectively.  For instance, {\bf 4} labels solutions with
$\umx=\infty$
        (i.e., $k\geq 0,\ h_a \geq 0,\ u_a>0$); {\bf 12} corresponds to the
        special case when the first zeros of $s_1$ and $s_2$ coincide and are
        smaller than those of $s_0$ and $s_3$ (if the latter exist), etc.

     Let us briefly outline the properties of the solution.
\begin{description}
\item[{\bf 0.}]
     The value $u=\umx =\pi /|k|$ corresponds to the second spatial infinity:
     $r^2\equiv \g_{\theta\theta}\to\infty$. The metric is regular; moreover,
     the asymptotic $u\to\umx$ is flat as well as $u\to 0$ (the proper
     asymptotic radius/circumference relation is valid). Thus the physical
     space section of the space-time forms a wormhole with (in general)
     different values of $\g_{tt}\equiv\e^{2\omega},\
     \g_{vv}\equiv\eta_v\e^{2\psi}$ and the scale factors $\e^{b_i u}$ at the
     two asymptotics.
\item[{\bf 1.}]
     $u=\umx$ is a singular sphere of a finite radius where
     $\g_{tt}=\e^{2\omega}\to\infty$.
\item[{\bf 2.}]
     The same as {\bf 1} with $\g_{tt}$ replaced by $\g_{vv}=
        \eta_v\e^{2\psi}$.
     In this case the 4-dimensional part of the metric is regular, so that
     the $D$-curvature singularity is connected with the 5th dimension.
\item[{\bf 3.}]
     At $u=\umx$ the radius $r$ is zero. The space-time has a singular
     center.
\item[{\bf 01.}]
     The same geometry as in the case {\bf 0} (the 3-dimensional section
     forms a wormhole) but $\g_{tt}\to\infty$ at $u\to\umx$.
\item[{\bf 02.}]
     The same as {\bf 01} with $\g_{tt}$ replaced by $\g_{vv}$.
\item[{\bf 03.}]
     $u=\umx$ corresponds to a finite radius; moreover, the integral
     $\int\sqrt{|\g_{uu}|}\,du$ diverges, meaning that the 3-dimensional
     space section forms an infinitely long tube, or ``horn'' like that
     described e.g. in \cite{banks}. The solution as a whole is nonsingular.
\item[{\bf 12.}]
     The same as {\bf 1} and {\bf 2} but both $\g_{tt}$ and $\g_{vv}$ are
     infinite at $u=\umx$.
\item[{\bf 13,\ 23.}]
     Space-times with a singular center where, respectively, $\g_{tt}$ or
     $\g_{vv}$ is infinite.
\item[{\bf 012, 013, 023, 123, 0123.}]
     The triple and quad\-ru\-ple combinations are also easily described, thus,
     in each case the figure {\bf 1} involved corresponds to a singularity of
     $\g_{tt}$, {\bf 2} to that of $\g_{vv}$ and the combination {\bf 03} to
     a ``horn''.
\end{description}

     Evidently the cases {\bf 0, 1, 2, 3} are general, the double
     combinations require an additional relation among the integration
     constants, the triple and quadruple ones are still more special.

     The possibility {\bf 0} can be realized only if all $h_a<0$, as follows
     from (\ref{IntGS}). This in turn means that all the charges $q,\ q',\
     \qq$ must be nonzero.

     On the other hand, the possibility {\bf 2} requires $h_2<0$ or/and
     $u_2<0$. This is possible only if $\eta_v=+1$. In the case $\eta_v=-1$
     the function $s(h_2,u+u_2)$ is replaced by the positive-definite
     function $h_2^{-1}\cosh h_2(u+u_2)$, with $h_2>0$. Along with the
     aforesaid, that means that all the variants involving {\bf 0} and {\bf
     2} are eliminated for configurations where the fifth coordinate $v$ is
     spacelike: from the above diversity of behaviors only the singular
     variants {\bf 1, 3} and {\bf 13} survive.

     In all cases when $\umx<\infty$ the ex\-t\-ra-di\-men\-si\-on scale
factors
     $\e^{b_i u}$ are regular in the whole space.

\begin{description}
\item[{\bf 4}.]
     $\umx=\infty$ if all $h_a>0,\ u_a>0$, whence by (\ref{IntGS}) $k\geq 0$.
     At the asymptotic $u\to\infty$ the metric coefficients $\g_{tt}\to 0$
     (unless $q=0,\ h_1=0$) and $\g_{vv}\to 0$ (unless $q'=0,\ h_2=0$).
     (Recall that at $q=0$ the function $s_1$ is replaced by $\e^{h_1 u}$ and
     similarly for $q'=0$ and $\qq=0$.) The behavior of the radius
     $r=\sqrt{|\g_{\theta\theta}|}$ depends on the relation between $k$ and
     $h_3$: $r\sim \e^{(h_3-k)u}$. In general, the surface $u=\infty$ is
     singular, with an exception deserving a separate description:
        solutions with BH and T-horizons.
\end{description}\nx
\subsection{Dyon black holes}
     As follows from Theorem 1 and is directly verified for
     (\ref{DsGS}), the most general solution with a BH horizon is
     (\ref{Dyon}), with three free parameters $k,\ q$ and
     $\qq$. The solution looks more transparent in curvature coordinates,
     with the notations
\begin{eqnarray}                                             
\nqq \al\al   r=R+\pp; \ \quad r_e=p=\sqrt{k^2+2q^2}-k;\nn
\nqq \al\al      r_m=\pp=\sqrt{k^2+2\qq^2}-k;\ \quad
               r_+=2k + r_m.                                \label{RR}
\end{eqnarray}
     Namely, the string metric and the dilaton field are
\begin{eqnarray}                                                     
     \ds \al=\al \frac{(r{-}r_+)(r{-}r_m)}{(r{+}r_e {-}r_m)^2}dt^2    \nn
   \al\al\nqq   -\frac{r^2 dr^2}{(r{-}r_+)(r{-}r_m)} -r^2 d\Omega^2
               +\sumi ds_i^2,                              \label{BHS}\\
\nqq \e^{2\lambda\varphi}\al=\al\biggl(1 +
\frac{r_e{-}r_m}{r}\biggr)^{2/(D-2)}.
                                                       \label{PhiBHS}
\end{eqnarray}
     In (\ref{BHS}) the extra dimensions are ``frozen'' and exert no
     influence on the 4-dimensional part of the metric, which is thus
     universal for all $D\geq 4$. The only trace of multidimensionality is
     the exponent in (\ref{PhiBHS}). Another observation of interest is the
     striking asymmetry between the electric and magnetic fields represented
     here by the parameters $r_e$ and $r_m$. This distinguishes dilaton field
     theory from the 4-dimensional Einstein-Maxwell theory.

     The space-time structure described by (\ref{BHS}) depends on the values
     of the three parameters $r_e\geq 0,\ r_m\geq 0$ and $r_+\geq r_m$:
\begin{description}
\item[(a)]
     $r_e=r_m=0$, corresponding to $q=\qq=0$: the Schwarzschild metric.
\item[(b)]
     $r_e=0,\ r_+=r_m>0:\ \g_{tt}\equiv 1;\ r=r_+$ corresponds to an
     infinitely long regular ``horn'', just the case described in
     \cite{banks} and papers cited therein.
\item[(c)]
     $r_e=0,\ r_+ >r_m>0$: the sphere $r=r_+$ is a Schwarzschild-like
     horizon, $r=r_m$ is a singularity inside it, with $\g_{tt}\to\infty$.
\item[(d)]
     $r_e>0,\ r_m=r_+ =0:\ \ r=0$ is a naked singularity, with $\g_{tt}\to
     0$.
\item[(e)]
     $r_e>0,\ r_m=0,\ r_+ >0:\ \ r=r_+$ is a Schwarzschild-like horizon,
     $r=0$ is a central singularity, again with $\g_{tt}\to 0$.
\item[(f)]
     $r_+ = r_m >r_e >0:\ \ r=r_+$ is a horizon of extreme RN type,
     $r=r_m-r_e$ is a singular sphere inside it ($\g_{tt}\to\infty$).
\item[(g)]
     $r_+ > r_m >r_e >0:\ \ r=r_+$ and $r=r_m$ are analogs of the outer and
     inner RN horizons; the sphere $r=r_m-r_e$ is like that in (f).
\item[(h)]
     $r_e\geq r_m>0;\ r_+=r_m:\ \ r=r_+$ is an extreme RN-like horizon,
     $r=0$ is a central singularity, where $\g_{tt}$ is infinite if $r_e=r_m$
     and finite if $r_e > r_m$.
\item[(i)]
     $r_e\geq r_m>0;\ r_+>r_m:\ \ r=r_+$ and $r=r_m$ are outer and inner
     RN-like horizons; $r=0$ is like that in (h).
\end{description}

     The dilaton field $\varphi$ is regular at all horizons, including inner
     ones, and singular at the singularities $r=0$ or $r=r_m-r_e$. In the case
     $r_m=r_e$, corresponding to $q=\qq$, the $\varphi$ field is constant and
     the 4-metric is just the RN one; in its usual notation
     [$\g_{tt}=(-\g_{rr})^{-1}=r^{-2}(r^2-2Mr+Q^2)$] its parameters $M$ and
     $Q$ are connected with ours as follows:
$$
     Q^2= 2q^2= q^2+\qq^2;\qquad   M=\sqrt{k^2+2q^2}\geq |Q|.
$$

     We see that, although the metric (\ref{BHS}) was obtained in search for
     solutions with horizons, one of its special cases (d) has a naked
     singularity, while another one, (b), is nonsingular and, in the opinion
     of some researchers, may describe the final state of evaporated BHs. To
     this end it should be emphasized that these ``horned particles'' form a
     very special subset in the set of solutions, with a single parameter,
     the magnetic charge.

\subsection{T-holes}
     In addition to the above family of BH solutions, there is a similar
     family of T-hole ones, obtained from the former by the substitution
     $t\chg v,\ q\to q'$, so that the horizons occur in the $(u,v)$ subspace.
     However, there are certain problems connected with the compactification
     of extra dimensions, which are most clearly understood on the following
        simple example.

        Putting $q'=0$ in (\ref{TH}), we come to a direct analog of the
        Schwarzschild solution (to be called {\it T-Schwarzschild}) which for
        $D=5$ coincides with the zero dipole moment soliton in the terminology
        of \cite{groper}:
\begin{eqnarray}                                                    
     && \nqq ds_D^2 = (1-2k/R)\eta_v dv^2 + dt^2 \nn
 &&\nqq -(1-2k/R)^{-1}dR^2 -R^2 d\Omega_2^2 +\sum_{i=2}^{n}ds_i^2 \label{TSch}
\end{eqnarray}
     while both fields $\varphi$ and $F$ are zero, so that
        $\e^{\lambda\varphi}\equiv 1$ and $\g_{AB}=g_{AB}$.

     At $R=2k$ the signs of $g_{uu}$ and $g_{vv}$ simultaneously
     change. Moreover, if $\eta_v =1$, i.e., this direction is
     timelike at big $R$, the overall signature of $V^D$ is preserved but in
     the opposite case, $\eta_v =-1$, it is changed by four: two spacelike
     directions become timelike. However, as is directly verified,
        a T-horizon is not a curvature singularity, either for the
        $D$-dimensional metric or for its 4-dimensional section.

     If $\eta_v=1$, the surface $R=2k$ is a Schwa\-rz\-sch\-ild-like horizon in
the
     $(R,v)$ subspace and an analytic continuation to $R<2k$ with the
        corresponding Kruskal picture exists.
        However, if some points on the $v$ axis are
     identified, as should be done to compactify $V_1$,
     then the corresponding sectors (wedges) are cut out in the Kruskal
        diagram, so that the $T$-domain and $R$-domain sectors join each other
        only in a single point, namely, the horizon intersection point.

     Another thing happens if $\eta_v=-1$. A further study is again possible
        after a transition to coordinates in which the metric is manifestly
        nonsingular at $R=2k$. Let us perform it for (\ref{TSch}) in the
        vicinity of the T-horizon, $R\to 2k$ (the more general case is treated
        similarly):
\begin{eqnarray}                                                    
        &&\nqq R-2k=x^2+y^2/(8k);\quad v=4k\arctan(y/x); \nn
        &&\nqq ds^2_2 (R,v) \approx
        \frac{R-2k}{2k}dv^2 + \frac{2k}{R-2k}dR^2
        = dx^2 + dy^2.                                  \label{Thor}\nn
\end{eqnarray}
        Thus the $(R,v)$ surface metric is locally flat near the T-horizon
        $R=2k$, transformed to the origin $x=y=0$, while the $v$
        coordinate behaves like an angle.

        This transformation could be conducted as a conformal mapping of
        the complex plane with the aid of the analytic function $\ln z,\
        z=x+ {\rm i}y$, as was done in Ref.\cite{br79} for some cylindrically
        symmetric Einstein-Maxwell solutions; then $v$ is proportional to
        arg$\,z$.

        Consequently, in the general case the $(R,v)$ surface near $R=2k$
        behaves like the Riemann surface having a finite or infinite (if $v$
        varies in an infinite range) number of sheets, with a branching point
at
        $x=y=0$ (a branching-point singularity \cite{br79}). If $V_1$ is
        compactified, $v$ is naturally described as an angular coordinate
     ($0\leq v < 2\pi l$, where $v=0$ and $v=2\pi l$ are identified and
     $l$ is the compactification radius at the asymptotic $R\to\infty$).
        $R=2k$ is just the center of symmetry in the $(R,v)$ surface; the
     latter has the shape of a tube having a constant thickness at
     $R\to\infty$, becoming narrower at smaller $R$ and ending at $R=2k$
     either smoothly (if the regular center condition $l=2(2k+p)$ is
     satisfied), or with a conic or branching-point singularity (otherwise).
        This suggests that there is no way to go beyond $R-2k$.

        In the singular case the geodesic completeness requirement is violated
        at the horizon, so it is reasonable to require $l=2(2k+p)$, or, more
        generally, $l=2j(2k+p)$ where $j$ is a positive integer, so that $R=2k$
        is a $j$-fold branching point. In this case a radial geodesic,
        whose projection to the ($R,v$) surface hits the point $R=2k$, passes
        through it and returns to greater values of $R$ but with another value
        of $v$, thus leaving the particular 4-dimensional section of the
        D-dimensional space-time. However, if the quantum wave function of the
        corresponding particle is $v$-independent, the particle does not
        disappear from an observer's sight and can look as if reflected from a
        mirror. It can be concluded that a T-horizon with $\eta_v=-1$ looks
        observationally like a mirror.

        Another thinkable possibility is to consider a continuation beyond
        $R=2k$, similarly to the way a cone is continued through its vertex. In
        this case, however, the space-time as a whole is no longer a manifold
        and, moreover, the neighborhoods of points belonging to the horizon
must
        be specially defined to preserve the Hausdorff nature of the
space-time.
        Whether or not it is possible, is yet to be studied. Physically such a
        continuation would mean that a particle getting to $R=2k$ "has a
choice"
        either to return to greater $R$, or to penetrate to smaller $R$, to the
        domain with another signature. One can assume that its probabilistic
        behavior is describable in terms of quantum concepts.

        If such an exotic possibility is not considered, the T-horizons
        are regular, although peculiar, surfaces of the $D$-dimensional
        space-time.

        The 4-dimensional sections of T-hole space-times are curved but
        nonsingular in the range $R\geq 2k$; a possible continuation to $R<2k$
        is discuused above and depends on $\eta_v$ or, if $\eta_v=-1$, on
        additional assumptions.

\section{\nhq $D$-dimensional solutions from the 4-dimensional viewpoint}
\markboth{7. The 4-dimensional viewpoint}{7. The 4-dimensional viewpoint}
     A 4-dimensional version of (\ref{Action}) is obtained by integrating out
     the extra-dimension coordinates, so that (up to a constant factor and
     a divergence)
\begin{eqnarray}                                                          
\nqq\al\al     S =  \int d^4 x \sqrt{^4 g}
     \e^{\sigma}\Bigl[R^{(4)}-\sigma^{,\mu}\sigma_{,\mu}
     + \sumi N_i\beta_{i,\mu}\beta_i{}^{,\mu} \nn
\nqq\al\al + \varphi^{,\mu}\varphi_{,\mu}
     - \e^{2\lambda\varphi} F^{\mu\nu}F_{\mu\nu}
        -2\eta_v\e^{-2\nu+2\lambda\varphi}W^{,\mu}W_{,\mu} \Bigr]  \label{Act4}
\end{eqnarray}
     where Greek indices range from 0 to 3, $^4 R$ is
     derived from the 4-dimensional part $g_{\mu\nu}$ of $g_{MN}$,
     and, as before, $\sigma= \sumi N_i\beta_i$.
     The Maxwell tensor $F_{\mu\nu}$ includes both $F^1$ and $F^3$,
     i.e., the electric and magnetic fields, while $F^2$ is re-formulated
     in terms of an effective scalar field $W$:
     $F^2= dW^2,\ W^2 = W(x^{\alpha})\,dv$ where, as before, $v$
pa\-ra\-met\-ri\-zes
     the one-dimensional subspace $V_1$ and $\nu=\beta_1$.
     Both $F_{\mu\nu}$ and $W(x^\alpha$)
     are coupled to $\varphi$ and all $\beta_i$. The field $W$ is minimally
     coupled to $g_{\mu\nu}$ and the sign of its kinetic term depends on
     $\eta_v$: it is normal if $v$ is spacelike and anomalous if $v$ is
     timelike.

     Eq.(\ref{Act4}) is written in the original $D$-Einstein
     conformal gauge.  The 4-dimensional Einstein gauge with the metric
     $\gg_{\mu\nu}$ is obtained by the conformal mapping similar to that used
     by Dicke \cite{dicke} and Wagoner \cite{wag}
\begin{equation}                                             
     \gg_{\mu\nu}= \e^{\sigma}g_{\mu\nu}                      \label{Map4E}
\end{equation}
     after which the action takes the form
\begin{eqnarray}                                             
\nqq\nqq \al\al    S=\int d^4 x \sqrt{^4 \gg}
     \Bigl[^4 \overline{R}+ \half\sigma^{,\mu}\sigma_{,\mu}
     + \sumi N_i\beta_{i,\mu}\beta_i{}^{,\mu}               \nn
\nqq \al\al\nq {} + \varphi^{,\mu}\varphi_{,\mu}
     {-}\e^{\sigma+2\lambda\varphi} F^{\mu\nu}F_{\mu\nu}
     {-}2\eta_v\e^{-2\nu+2\lambda\varphi}W^{,\mu}W_{,\mu} \Bigr]  \label{Act4E}
\end{eqnarray}
     where indices are raised and lowered using $\gg_{\mu\nu}$.

     The actions (\ref{Action}) and (\ref{Act4E}) are equally convenient for
     solving the field equations due to the constant effective gravitational
     coupling. Noteworthy, the coordinate $u$ as introduced in
     (\ref{Harm}) is harmonic with respect to both the $D$-metric
     $g_{MN}$ and the 4-metric $\gg_{\mu\nu}$, i.e., $\nabla^M
     \nabla_M u = \overline{\nabla}^{\mu}\overline{\nabla}_{\mu} u =0$, but
     not with respect to the 4-dimensional part $g_{\mu\nu}$ of the
     $D$-metric:  $\nabla^{\mu}\nabla_{\mu} u \ne 0$.

     The metric $\gg_{\mu\nu}$ thus corresponds to the so-called
     {\sl gravitational system of measurement} \cite{dicke,stanmel}. However,
        real space-time measurements, such as solar-system experiments,
        rest on the constancy of atomic quantities (the {\sl atomic system of
        measurements}. Thus, the modern definition of reference length is
        connected with a certain spectral line, determined essentially by the
        Rydberg constant and ultimately by the electron and nucleon masses.
        Therefore observational properties of various theoretical models
        are most reasonably described in a
        conformal gauge where masses of bodies of nongravitational matter, such
        as atomic particles, do not change from point to point.

        In other words, in the gauge to be selected (to be denoted
        $g^*_{\mu\nu}$) the nongravitational matter Lagrangian $L_m$ should
        enter into the action with no $\sigma$- or $\varphi$-dependent factor.
        However, the choice of $g^*_{\mu\nu}$ depends on how $L_m$ appears in
        the original action, that is, how matter is coupled to the metric and
        dilaton fields in the underlying fundamental theory.

        In \cite{green}, where the effective field-theoretic limit of string
        theory in 10 dimensions is given in a form similar to (\ref{Action})
        (Eq.(13.1.49)), some quadratic fermion terms do not contain the
        dilaton.  If those terms are associated with matter, then in our
        simplified model it is reasonable to write $L_m$ just as an additional
        term in the brackets of Eq.(\ref{Action}). Then, passing over to the
        4-dimensional formulation, it is easy to check that the metric in the
        ``atomic gauge'' should have the form
        \begin{equation}                                          
        g^*_{\mu\nu}  =  \e^{\sigma/2}g_{\mu\nu}              \label{MapA}
        \end{equation}

        The term $L_m$ would enter (\ref{Act4}) and (\ref{Act4E}) with the
     factors $\e^{\sigma}$ and $\e^{-\sigma}$, respectively, whereas in
     terms of $g^*_{\mu\nu}$ the matter part of the action is just
        $\int d^4 x\sqrt{g^*} L_m$. The same metric $g^*_{\mu\nu}$ would be
     obtained if we wrote the action for a point particle moving in $D$
        dimensions in the conventional form $-\int m\, ds $ and required that
        it move along geodesics of the 4-dimensional metric $g^*_{\mu\nu}$.

        The notion of active gravitating mass of an isolated object in a space
        with the structure (\ref{Stru}) is to be also introduced with the aid
of
        $g^*_{\mu\nu}$, by comparing $g^*_{tt}$ far from the source with the
        expression $(1-2GM/r)$ of the Schwarzschild metric, so that, with an
        arbitrary radial coordinate $u$ \cite{br-acta}
\begin{equation}                                                   
     GM= -|g^*_{\theta\theta}|^{3/2}\frac{\partial_u g^*_{tt}}
              {\partial_u g^*_{\theta\theta}}\Bigg|_{u\to u_{\infty}}
     \nq  =  \frac{r^2{\gamma^*}'}{r'}\Bigg|_{u\to u_{\infty}}
                     \label{Mass}
\end{equation}
     where $\e^{2\gamma^*}=g^*_{tt},\
     r^2 = -g^*_{\theta\theta}$ and $u_{\infty}$ is the value of $u$ where
        $r\to\infty$ and $\gamma^*\to 0$.

\def\frD{\frac{1}{4(D-2)}}

        As in Sect.6, let us consider only solutions for $\lambda=\lst$. In the
        general case of Solution G the mass is expressed in terms of the
     constants $k,\ h_a,\ q_a,\ b_i$:
\begin{eqnarray}                                                      
        GM \al=\al \frD \bigl[ \sum_{a=1}^{3}c_a\sqrt{2q_a^2+h_a^2\sign h_a}
            +2b \bigr];   \nn
        \al\al\nq    c_1=3D-8,\quad   c_2=-2,\quad   c_3=D.          \label{MG}
\end{eqnarray}
        For the dyon solution (\ref{Dyon}) the mass is connected with the
        charges and the other constants in the following way:
\begin{eqnarray}
        GM= k + \frD \bigl[(3D-8)p + D\pp\bigr]\qquad          \nn
          = \frac{r_+}{2} + \frD \bigl[(3D-8)r_e + (D-4)r_m\bigr]
\label{MDyon}
\end{eqnarray}
        The expression (\ref{MDyon}) can be obtained from (\ref{MG}) by
        substituting $q'=h_2=b=0,\ h_1=h_3=k$. Noteworthy, the mass
(\ref{MDyon})
        is nonnegative, while in the general case (\ref{MG}) the sign is not
        fixed.

        For T-holes with a magnetic charge, obtained from (\ref{Dyon}) by the
        substitution $t\chg v,\ q\to q',\ p\to p'$, we obtain:
\begin{equation}
        GM= \frac{k}{4} + \frD (D\pp-2p').                       \label{MTH}
\end{equation}

        For other solutions $GM$ can be easily found as well using (\ref{MapA})
        and (\ref{Mass}).

        Eq.(\ref{MDyon}) generalizes the corresponding relations for dilatonic
        BHs in 4 dimensions and Eq. (67) of Ref.\cite{br10} for electrically
        charged BHs in $D$ dimensions. Thus, extreme BHs, those with
        the smallest mass for given charges, correspond to $k=0$.  At $k\to 0$
        the horizon is squeezed to a point (the center) and becomes a
        singularity.

        As for T-holes, Eq.(\ref{MTH}) (generalizing Eq.(68) from
        \cite{br10}) shows that they can have negative
        gravitational masses, i.e., be felt by test particles as repellers.
That
        happens if $\eta_v = +1$ for large $q'$ (see (\ref{TH}).

     In conformal gauges other than (\ref{MapA}) test particle trajectories
        $x^j (t)\ (j=1,2,3)$ are certianly the same but they are no longer
        geodesics: particles fell both the metric and the scalar
        field. A similar situation was discussed by Dicke \cite{dicke}.

\section{Discussion}                                     
\markboth{8. Discussion}{8. Discussion}
\noindent {\bf 8.1.} The solutions obtained here are less general as compared
        with those of Refs. \cite{galz,myers94} in that they do not contain the
        Taub-NUT parameter, axion charge and Kerr rotation parameter. However,
        they are more general in that they include extra dimensions (the
        corresponding $n$ parameters are $b_i$ and the additional gauge field
        component $F^2$ (the parameter $q_2=q'$).

        Another point of interest is
        connected with the possible non-string couplings: the remarkable
        symmetries of string theory, which enabled the authors of Refs.
        \cite{galz,myers94} and some earlier papers to obtain string-theory
        solutions without actually solving the equations, do not work when
        $\lambda\ne\lst$.  Some such solutions are presented here but they are
        less general than those with $\lambda=\lst$. Apparently the same
        symmetries explain the total integrability of the field equations
        which follow from string theory.

        The approach connected with a direct study of the field equations has
        made it possible not only to obtain some solutions within less
symmetric
        versions of field theory, but also to prove certain statements
(Theorems
        1 and 2) for situations when it is hard to obtain exact solutions.
\medskip

\noindent {\bf 8.2.}
        The stability of Solution E under small perturbations preserving
        spherical symmetry was studied in Refs.
        \cite{br-vuz92,bm-itogi,potsdam}. The system with two dynamic degrees
of
        freedom (the dilaton field and a single extra-dimension scale factor)
        was considered and three cases when the perturbation equations decouple
        were studied in detail. It was concluded that solutions with naked
        singularities are catastrophically unstable, while the BH ones are
     stable. These results generalized the earlier ones from \cite{brohod} and
     \cite{brokir} where the instability of static scalar-(electro)vacuum
     configurations in conventional general relativity, in
     particular, black holes with scalar charge, was established.

     In Ref. \cite{br10} the earlier results were confirmed and
        slightly extended; in particular, the simplest example of a T-hole
        (the uncharged one, Eq.(\ref{TSch}) with $a_0=0,\ a_1=1$) was
        investigated and shown to be unstable.

     Among the \sss configurations in dilaton gravity studied to-date only BHs
        turned out to be stable under monopole perturbations. It would be of
        importance to extend the stability investigation to other
        configurations, in particular, those described here and in the papers
        \cite{galz,myers94}.
\medskip

\noindent {\bf 8.3.}
     We have seen that T-holes, the possible windows to
     space-time domains with unusual physics, appear as solutions to
     multidimensional field equations as frequently as do black holes.

     As in the present state of the Universe extra dimensions are generally
     beleived to be compactified to a very small size, it is hard to imagine
     how T-holes might now form from ordinary matter. However, in the early
     Universe where all dimensions could be equally relevant, T-holes could
     form on equal grounds with primordial black holes and consequently their
     relics might play a certain role at the present stage, e.g., be one
     of the forms of dark matter (as suggested, e.g. by Wesson \cite{wesson}.
        However, such a possibility looks questionable due to the instability
of
        these objects. The latter is so far established only for the
        above simplest, uncharged T-hole solution.
        However, very probably the same is true for charged T-holes
     since perturbations near a T-horizon behave just as they do near a
     singularity \cite{br10}, while all singular solutions studied so far
     \cite{br-vuz92,bm-itogi,potsdam} turned out to be unstable.
     Nevertheless, T-hole solutions in other field models, which
     exist for sure, may turn out to be stable, although, on the
     other hand, it may happen that there is a kind of ``censorship'' like
     Hawking's chronology protection conjecture \cite{hawk}, confirmed, in
        particular, by the instability of Cauchy horizons.
\medskip

\noindent {\bf 8.4.}
     Certain difficulties in the T-hole description arise due to the
     compactification of extra dimensions. However, the latter may be
     invisible to 3-dimensional observers for a reason other than their small
     size, such as, e.g., the behavior of field potentials, as discussed in
     Ref. \cite{rubshap} (we may live within a 3-dimensional ``membrane'' at
     the bottom of a potential ``trench'' in a multidimensional world).
     A possible BH and T-hole existence in such models may be a subject of
        interest for further study.

\Acknow{This work was in part supported by the Russian Ministry of Science.
        I would like to thank my colleagues D.Galtsov, V.Ivashchuk,
        M.Konstantinov and V.Melnikov for useful discussions.}

\end{document}